\documentclass[12pt]{article}
\usepackage{latexsym,amsmath,amssymb,amsbsy,graphicx}
\usepackage[utf8]{inputenc}
\usepackage[T2A]{fontenc}
\usepackage[space]{cite} 
\usepackage{multirow}

\usepackage{graphicx}
\usepackage{dcolumn}
\usepackage{bm}
\usepackage{todonotes}

\makeatletter\def\@biblabel#1{\hfill#1.}\makeatother

\begin{document}

\noindent\begin{minipage}{\textwidth}
\begin{center}

{\large The $BVR_cI_c$ sky brightness of the Caucasian Mountain Observatory of MSU}

Komarova I.\,A.$^{1,2}$, Tatarnikov A.\,M.$^{1,3a}$, Sharonova A.\,V.$^2$, Belinskii A.\,A.$^3$, Maslennikova N.\,A.$^3$, Ikonnikova N.\,P.$^3$, Burlak N.\,A.$^3$\\[6pt]

\textit{$^1$Lomonosov Moscow State University, Faculty of Physics, Moscow, 119991 Russia}
\textit{$^2$Lomonosov Moscow State University, Faculty of Space Research, Moscow, 119991 Russia}
\textit{$^3$Sternberg Astronomical Institute, M.V.Lomonosov Moscow State University, Moscow 119191, Russia.}\\
\textit {E-mail: $^a$andrew@sai.msu.ru} \\ [1cc] 
\end{center}

{\parindent5mm In this paper we analyse the measurements of the brightness of the night sky above the CMO SAI MSU in the visible and near-infrared range made in 2019-2014. In 2023-2024 the median zenith brightness of the moonless night sky was $21.31^m$ in the $B$ band, $20.63^m$ in the $V$ band, $20.15^m$ in the $R_c$ band, and $19.11^m$ in the $I_c$ band. In 5 years the sky brightness had increased by $0.7^m$ in $B$ and $V$, by $0.45^m$ in $R_c$, and $~\sim0.1^m$ in $I_c$. We found that the brightness growth is mostly (up to $\sim85$\%) due to the increasing light pollution from nearby cities, while the remainder can be attributed to the increase of solar activity after the 2019 minimum. We discuss how the sky brightness is influenced by such factors as the airmass and the location of the Sun and the Moon in the sky. A qualitative analysis of the sky emission spectrum has demonstrated the growing role of LED lamps in light pollution. These changes in sky brightness which are only going to get harder favour observations that are less sensitive to the degree of light pollution - IR photometry and spectroscopy and high-resolution optical spectroscopy.
\vspace{2pt}\par}

\vspace{1pt}\par

\vspace{1pt}\par
\end{minipage}

\section*{Introduction}

The Caucasian Mountain Observatory of the Sternberg Astronomical Institute (CMO SAI MSU) is a high-altitude observatory (>2000~m above sea level) located in the North Caucasus on Mount Shatdzhatmaz 20 km south of Kislovodsk. At present, two telescopes of the observatory (\cite{Shatsky2020}) with mirrors of 0.6-m and 2.5-m are actively engaged in photometric and spectroscopic observations in the optical and infrared ranges requested by the scientists of the Lomonosov Moscow State University, INASAN, SAO, Ural Federal University, etc.

The astroclimate and meteorological conditions at the CMO site are well investigated, and the corresponding measurements in the optical range have been carried out since 2007 to the present. The results obtained up to 2015 were published in \cite{Kornilov2014}, \cite{Kornilov2016a} and \cite{Kornilov2016b}. They show that the observatory is one of the best in Russia in terms of parameters such as the image quality and amount of precipitable water (PWV) in months with the highest amount of clear night sky. Despite the proximity of the city, CMO also had a fairly good score in terms of sky brightness. Thus, according to \cite{Kornilov2016a}, the surface sky brightness in the $B$ band on moonless nights in 2007~-- 2011~ was $22.1^m$/arcsec$^2$ (median value) or 22.3$^m$/arcsec$^2$ (first quartile of the distribution). This is typical of observatories located at altitudes less than 3~km, whereas the observatories located far from urban light at altitudes greater than 3~km are characterised by darker skies with $B = 22.8^m$/arcsec$^2$ (\cite{Pedani2009}). 

Recently, the night-sky brightening due to the accelerating urbanisation process and the development of artificial lighting has become an increasingly urgent problem for many observatories, including those located in mountainous areas. Meanwhile, this very parameter of the astroclimate largely determines the limiting brightness of objects available for observations. The signal-to-noise ratio in aperture photometry (neglecting the dark current noise) is expressed as follows:

\begin{equation}
SNR =\frac{F_0 \cdot t }{\sqrt{F_0 \cdot t + n\cdot s \cdot F_{bg} \cdot t + n \cdot RN^2}},
\label{eq:snr}
\end{equation}

where $t$ is the exposure time, $n$ is the number of pixels in the star image, $s$~--- the sky area corresponding to 1 pixel (in arcsec$^2$), $RN$ is the readout noise, and $F_0$ and $F_{bg}$ are photon fluxes from the object and from 1 arcsec$^2$ of the sky, respectively. 

The mentioned above characteristic sky brightness in the CMO for the 2.5-m telescope corresponds to a rather large value of $F_{bg}$~~---approximately 60 ph$\cdot$arcsec$^{-2}$s$^{-1}$.
Thus, for faint stars, the relation $F_0 \ll F_{bg}$ is fulfilled. Since large exposure times are used in observations of faint objects, we can neglect the readout noise, which is independent of time. In this case, the formula \ref{eq:snr} is transformed into a simple relation $SNR =\sqrt{t} \cdot F_0 / \sqrt{n \cdot F_{bg}}$, which shows that the increase of sky brightness by 0.75$^m$ (i.\,e. by a factor of 2) requires doubling the integration time to obtain the same SNR.

The sky brightness is composed of several components: 1) extra-atmospheric background emission consisting of zodiacal light and radiation from faint unresolved objects, 2) airglow which is the emission of light by the upper layers of the Earth's atmosphere due to ion recombination, luminescence and chemiluminescence, 3) instrumental background and 4) light pollution from infrastructure facilities. While the former two components are relatively constant (the value of the atmosphere's own luminescence can be influenced by solar activity), and the instrumental background in the optical range can be neglected, the influence of the anthropogenic factor is, firstly, seasonal, and secondly, increasing with time. In the case of the CMO, the neighbour producing light pollution is the Caucasian Mineral Waters resort, located to the north of the observatory, which is now actively developing. In addition, since 2020, bright night lighting has been introduced along the Kislovodsk~--- ‘Valley of Narzans’ highway, which runs in close proximity to the observatory.

In \cite{Tatarnikov2024} we have investigated the brightness of the CMO sky in the near-infrared. We have shown that it is still at the level of the best world observatories and did not show a systematic increase during 2016~-- 2023. The aim of this work is to measure the current brightness of the CMO sky in the $B$, $V$, $R_c$,  $I_c$ photometric bands and to investigate its dependence on the main meteorological parameters and the Sun and Moon coordinates, as well as to determine the anthropogenic contribution to the total sky brightness using the data obtained with the 60-cm telescope from 2019 till 2024.

\section*{Observations}

The observations used in this work for measuring the sky brightness were carried out in the $BVR_cI_c$ bands from 2019 to 2024 on the RC600 telescope installed at the CMO of SAI MSU in 2019 (\cite{Berdnikov2020}). The detector was an Andor iKon-L DZ936N-BV 2048x2048 camera, with a pixel size of 13.5~$\mu$m, cooled to $-70^\circ$~C and having low dark current and low readout noise. The image scale is $49.1''$/mm or $0.66''$/pixel. Most of the observations were made in low-cloud or cloudless nights, with humidity less than 95~\% and wind speed below 10~m/s. The primary goal of observations was to study astronomical objects of various types~--- from exoplanets to active galactic nuclei. However, besides the program objects which occupy only a small fraction of the image, the background radiation was recorded in each frame, too. More than 10000 frames were obtained in each photometric band during this time.  

The spectra used in this work were obtained with the TDS instrument (\cite{Potanin2020}) of the 2.5-m CMO telescope on photometric nights. The exposure time of each frame was 10 min, and the spectra were recorded simultaneously in two channels, red and blue, with a boundary at about 5700\AA{}.

\section*{Calibration}

The total background signal measured on the frames obtained with an exposure $t_{exp}$ consists of a fixed offset (\textit{bias}), dark current ($dark \cdot t_{exp}$), and background sky ($sky \cdot t_{exp}$). This signal is contaminated by the noise of the signal components and readout noise ($RN$), which is independent of the exposure time. To reduce the influence of noise on the background brightness value we averaged the signal from a large number of pixels.

\begin{figure}[h]
\includegraphics[width=0.5\textwidth]{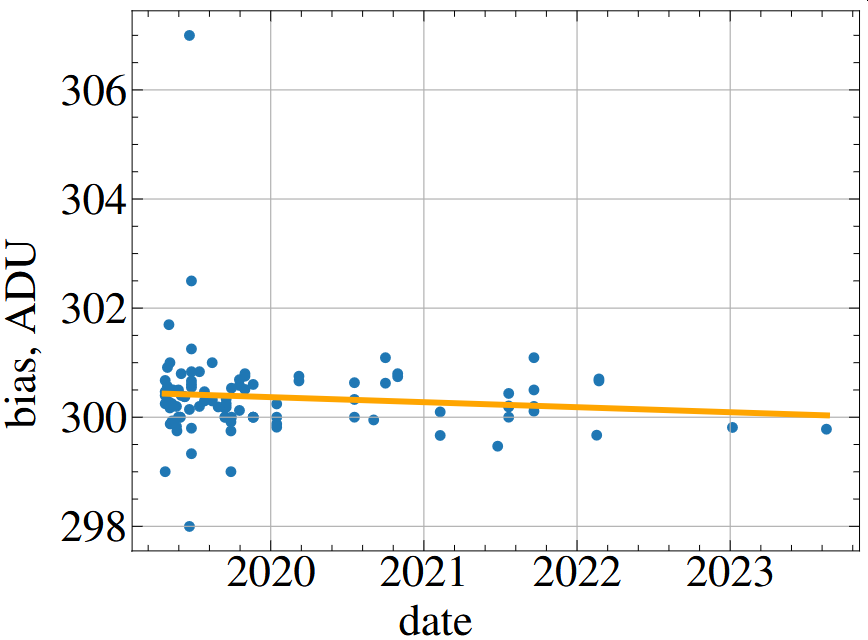} 
\caption{\textit{Bias} level of the RC600 camera in $2019-2023$} 
\label{fig:bias}
\end{figure}

The instability of the \textit{bias} level and the value of the dark current can directly affect the measured background value. The dark current of the Andor iKon-L DZ936N-BV camera operating at a temperature of $-70^\circ$~C is $\sim0.001$~e$^-$/s according to our measurements. This is orders of magnitude smaller than the expected sky brightness, so its variation can be neglected. The changes in the \textit{bias} level relate to each particular frame. So, the \textit{bias} cannot be measured directly on each science frame. Fig.~\ref{fig:bias} shows the change in the average $bias$ level over the entire observation period. It can be seen that the average value was 300.5 counts (or analogue-digital units, ADUs). There are deviations from this value which rarely exceed $\pm1$~ADU.

For measuring sky brightness, we selected the frames with integration time not less than 30~s for the $B$ and $V$ bands and not less than 10~s for the $R_c$ and $I_c$ bands where the sky brightness in ADU is higher. Since the value of \textit{bias} does not depend on the exposure, this restriction allows us to reduce the influence of \textit{bias} instability on the final result. \textit{Bias} and dark frames were subtracted from the selected science frames. The median signal in ADU was accepted as the value of sky brightness for each frame. This effectively removed the contribution of stars to the measured signal.

This processing step resulted in a table containing the date, observing conditions, photometric band and sky brightness in units of ADU/(arcsec$^{2}$\,s).

\begin{figure*}
\includegraphics[width=\textwidth]{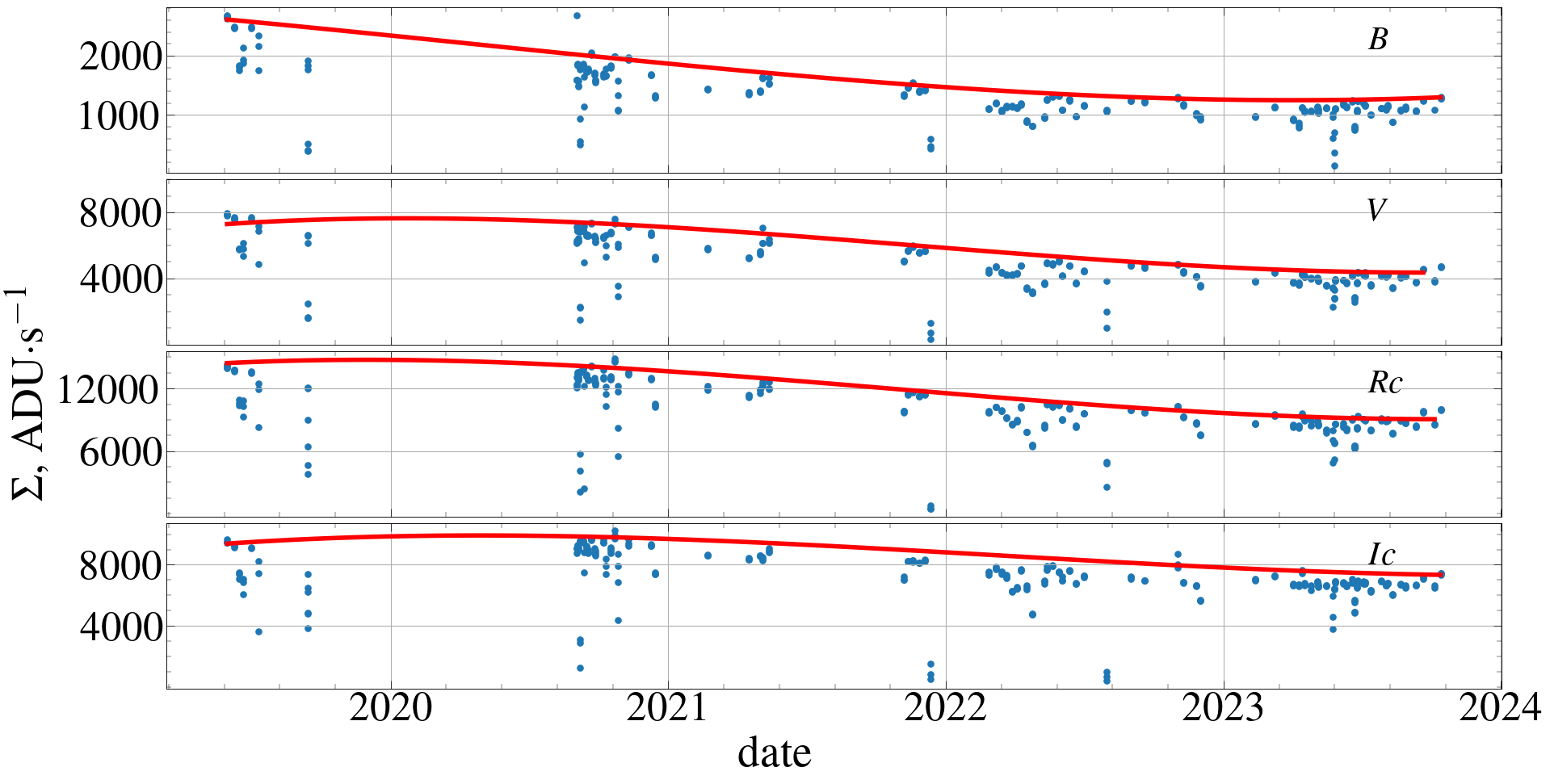} 
\caption{Variation of the number of counts (in ADU) registered from the standard star in the photometric bands $B$, $V$, $R_c$, $I_c$. The red line shows the upper envelope corresponding to observations on clear nights}. 
\label{fig:v1413}
\end{figure*}

To convert these units to magnitudes, we used the star located at R.A.\,=\,19:03:47, DEC\,=\,+16:26:28 and with the assigned magnitudes $B=13.938^m$, $V=13.134^m$, $R_c=12.642^m$, and $I_c=12.146^m$ which serves as the photometric standard for the variable star V1413~Aql regularly observed with RC600. This star is suitable in the sence that it is observed at the CMO within a relatively narrow range of airmasses. Figure ~\ref{fig:v1413} shows the variation in the number of counts detected from the standard star within wide photometric aperture (aperture radius $R_{ap}=20$ pixels, or nearly $13^{\prime\prime}$). One can see a general trend of decreasing flux associated with the degradation of the mirror reflectivity due to contamination. Similar results for the mirror reflectivity degradation are described in \cite{Boccas2004}. Besides the general trend, there are seen the measurements dropping down. These are associated with observations under poor atmospheric transparency conditions, so we used the upper envelope of the points to construct the calibration curve.

Hereinafter we consider the data corrected for varying transmission.

\section*{Sky brightness at different airmasses}

\begin{figure*}
\includegraphics[width=\textwidth]{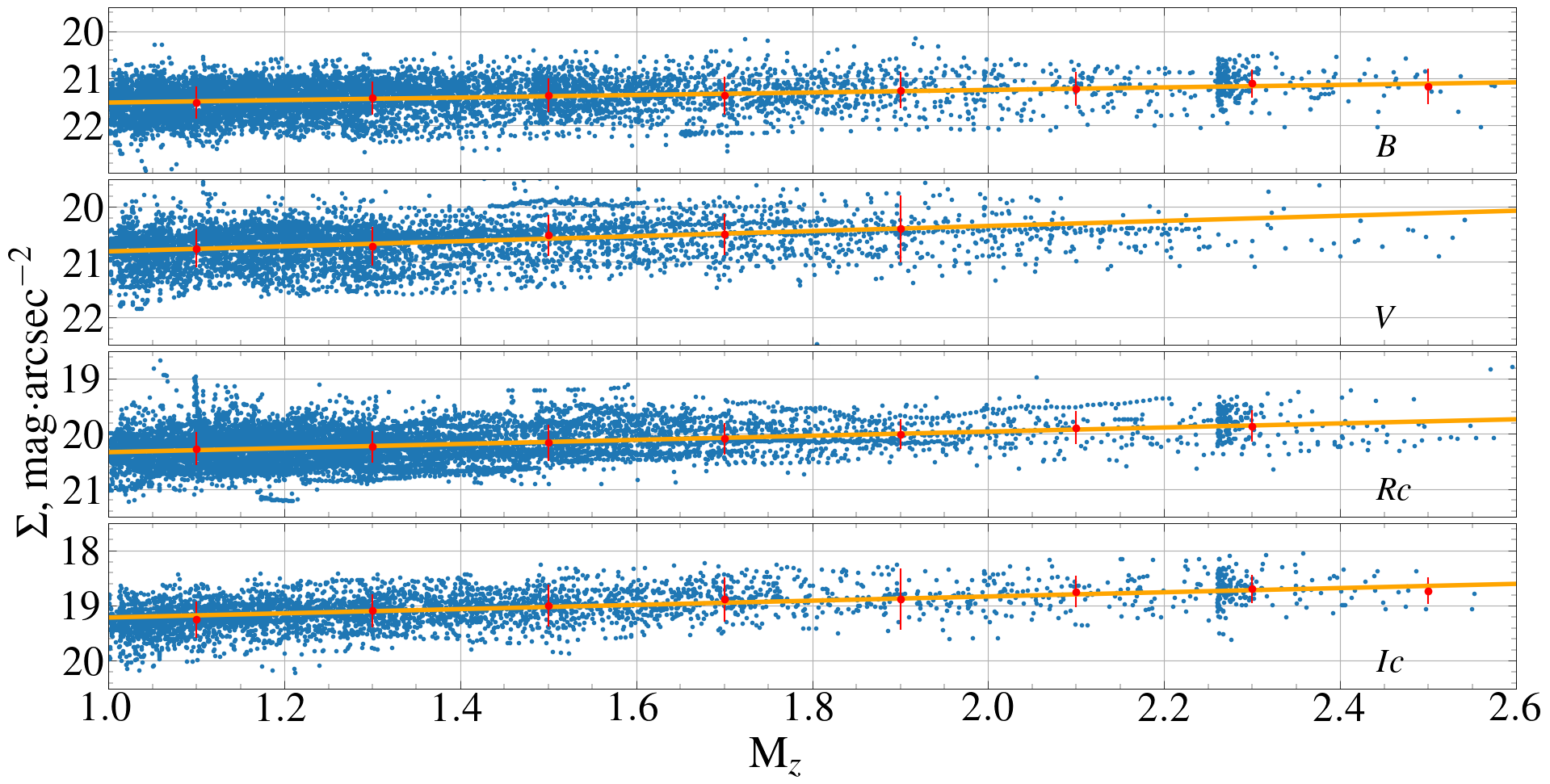} 
\caption{The dependence of the surface sky brightness $\Sigma$ on the airmass in the $B$, $V$, $R_c$, and $I_c$ bands for night time with the Moon being at phase $<0.5$ and below the horizon. The red dots are the sky brightness values averaged within the bins of $\Delta Mz=0.2$, the orange line runs through the red dots}.
\label{fig:bg(mz)}
\end{figure*}

Fig.~\ref{fig:bg(mz)} shows the dependence of the $BVR_cI_c$ surface sky brightness $\Sigma$ on the airmass based on the data obtained on moonless nights (i.e., when the Sun is more than $18^\circ$ below the horizon) during the entire observational period. The plots demonstrate a big scatter of points, associated, first of all, with the presence of a general trend of sky brightening (see below). The cluster of points at $Mz\approx2.3$ in the $B, R_c$ and $I_c$ bands is noteworthy. These points refer to one particular object with large negative declination, which can be observed at the CMO only near its upper culmination at an altitude of $\sim25^\circ$ (the $V$ frames were not selected for measuring because of short exposures which did not satisfy the criteria set at the beginning of the study). In addition, one can see the chains of points in the $V$ and $R_c$ bands~---these are the result of follow-up observations, which can last for hours, and when a large number of individual images with short exposures is obtained.

Despite the significant scatter of points, it can be seen that the mean values of surface brightness lie well on the straight lines. 
This is particularly clear when looking at follow-up observations of the same region of the sky the points in the $Mz - \Sigma$ diagram line up on average along the straight lines, demonstrating that the brightness (in magnitudes) grows uniformly towards the horizon.

\section*{Sky brightness changes over time}

\begin{figure*}
\includegraphics[width=\textwidth]{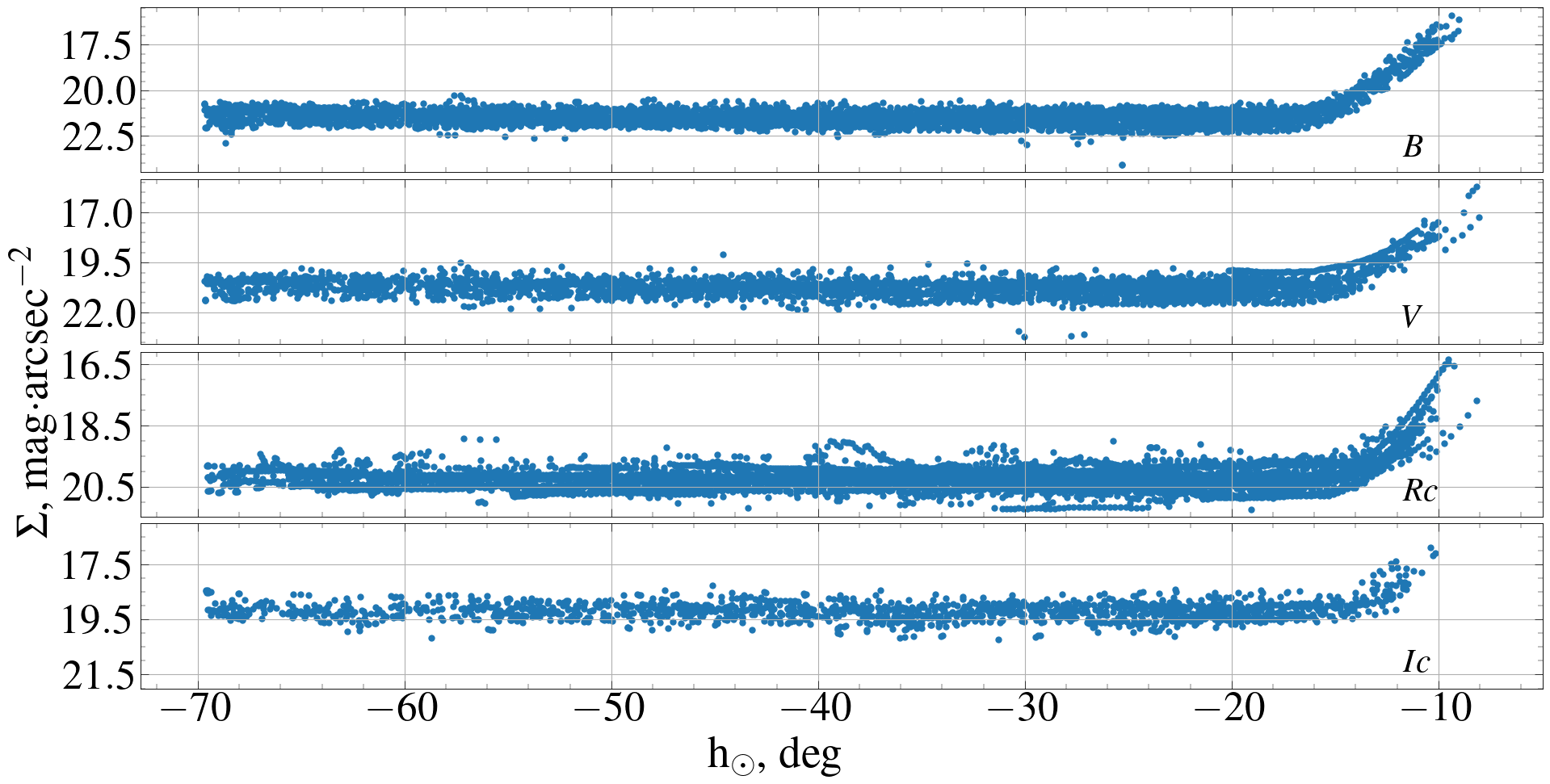} 
\caption{
Dependence of the surface sky brightness $\Sigma$ on the Sun's altitude for airmasses $Mz<1.5$ in the $B$, $V$, $R_c$, and $I_c$ bands. The Moon's altitude is less than $0^\circ$ with the phase being $<0.5$.
}
\label{fig:sun}
\end{figure*}

The brightness of the twilight and night sky strongly depends on the Sun's depression below the horizon. It is believed that the Sun ceases to affect the sky brightness after dipping below the horizon for more than $18^\circ$ (the end of astronomical twilight, the beginning of night). Fig.~\ref{fig:sun} shows the dependence of the sky brightness at small airmasses on the Sun's altitude in the absence of the Moon. As expected, the sky brightness in different bands reaches a plateau at different depths of the Sun below the horizon~--- in the IR band $I_c$ at a dip greater than $13^\circ$, in $R_c$ at a depth of $14^\circ$, in $V$ at a depth of $15^\circ$, and in the $B$ band at a dip of $16^\circ$. 

\begin{figure*}
\includegraphics[width=\textwidth]{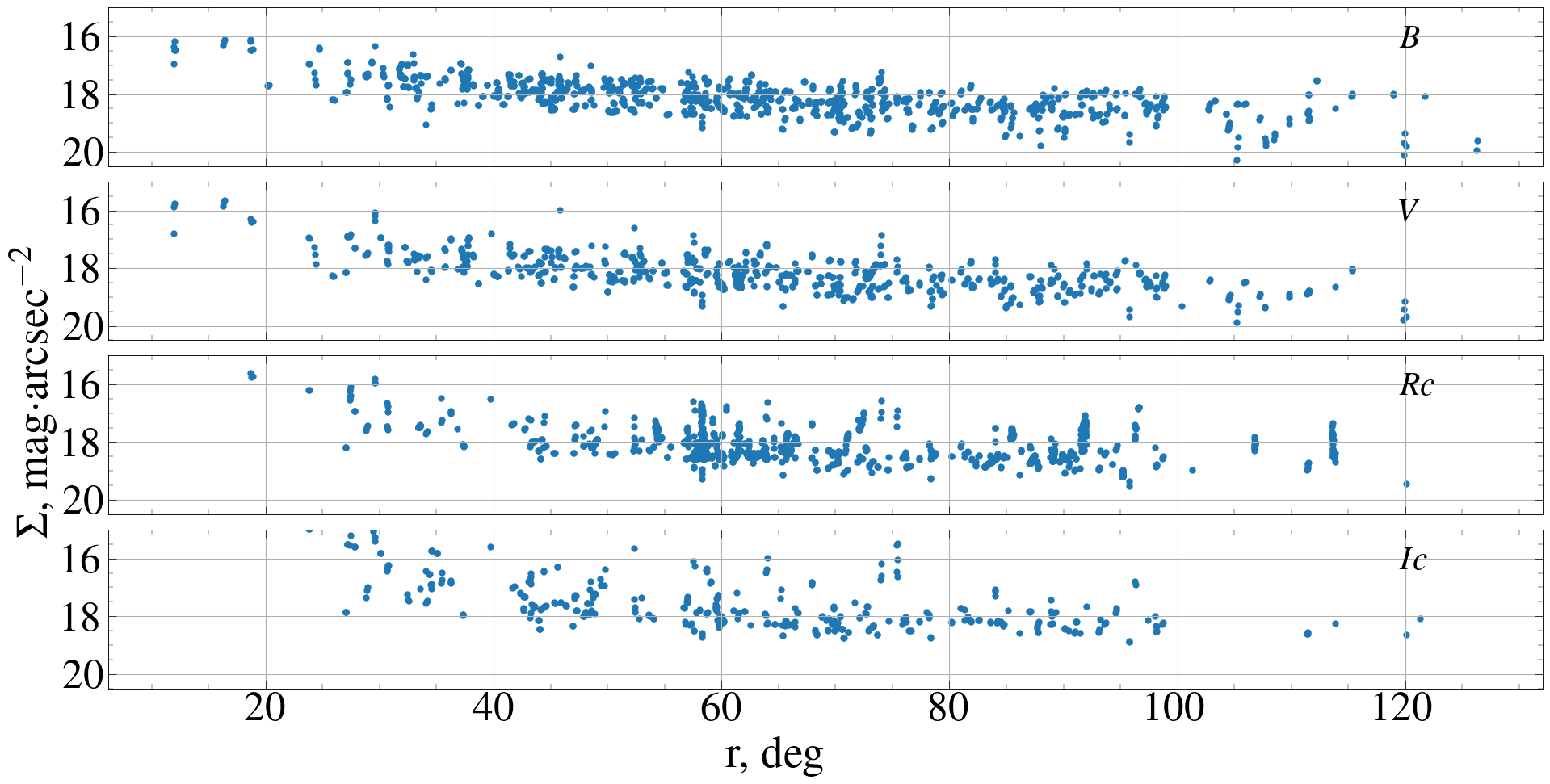} 
\caption{
Dependence of the sky brightness on the angular distance to the Moon for night time, $Mz<1.5$ and the Moon phase $\ge 0.97$.
}
\label{fig:moon}
\end{figure*}

In addition to the Sun, the Moon has a great influence on the sky brightness, too (especially near the full moon). We selected images taken at night high above the horizon when the Moon phase was $\ge 0.97$. Fig.~\ref{fig:moon} shows the dependences of the sky brightness on the distance to the Moon under the above specified conditions. It can be seen that in all bands the brightness decreases relatively slowly with distance to the Moon and even at a distance of several tens of degrees remains at $18^m-19^m$/arcsec$^2$, which corresponds to the sky brightness during navigational twilight (see Fig.~\ref{fig:sun}).

To investigate the long-term variability of sky brightness, we used measurements obtained at night, with the Moon height $<0^\circ$ and phase $<0.5$. Selecting not only by the Moon height but also by phase was aimed to exclude the lunar dawn effect~--- the increase in sky brightness before moonrise, which can be noticeable at large Moon phases.

\begin{figure*}
\includegraphics[width=\textwidth]{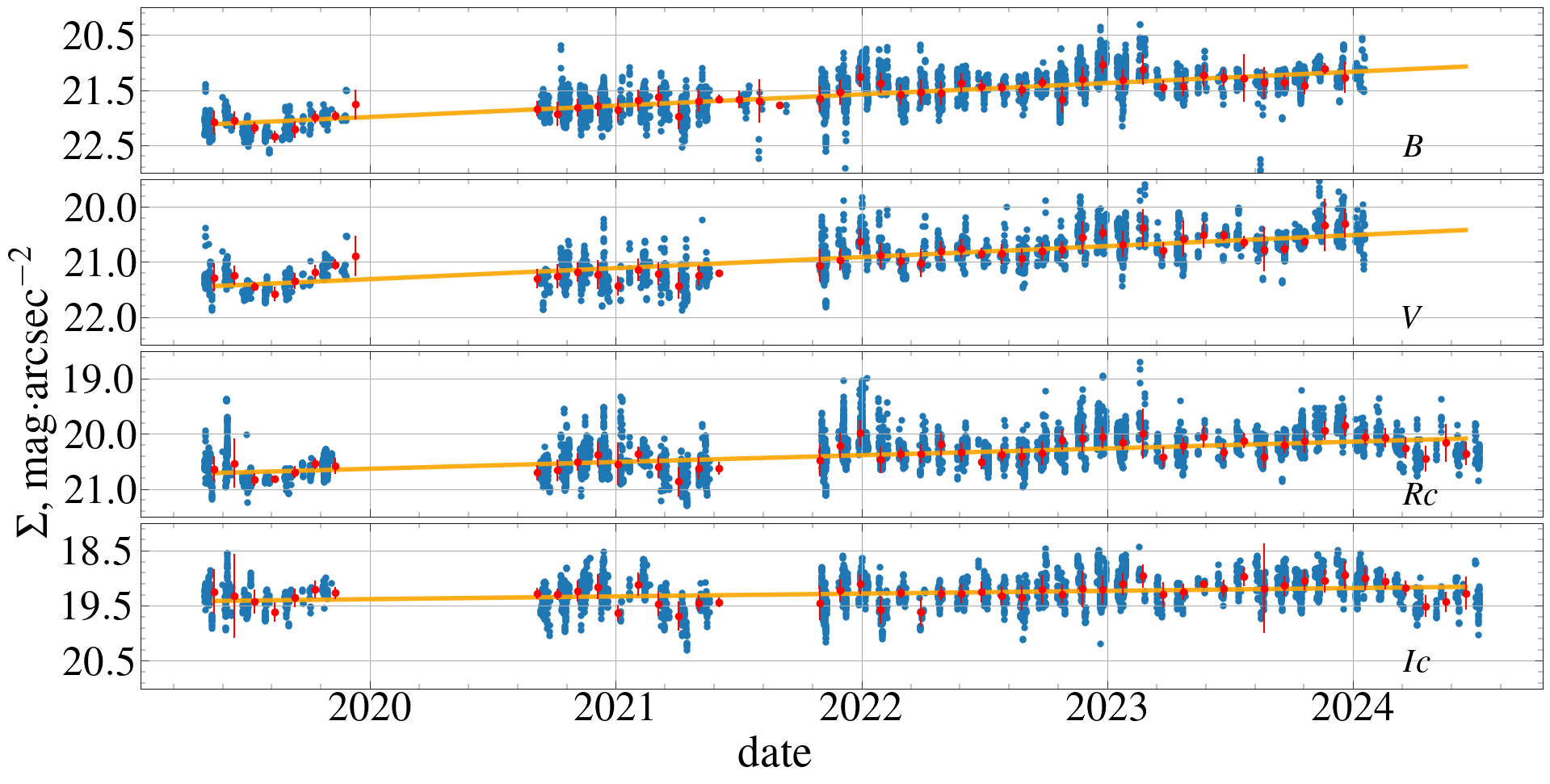} 
\caption{
Dependence of the surface sky brightness $\Sigma$ on time in the $B$, $V$, $R_c$, and $I_c$ bands for night time with the Moon being at phase $<0.5$ and below the horizon.
}
\label{fig:mag(time)}
\end{figure*}

Fig.~\ref{fig:mag(time)} shows the variation in surface brightness of the moonless night sky over the period from 2019 to 2024. Numerous short gaps in the data refer to full-Moon nights and long gaps in 2020 and mid-2021 -- to a software failure of the storage system. The plots also show seasonal variations in sky brightness and a general trend of increasing brightness with time. 

The seasonal variations are probably related to periodic snowfalls in the vicinity of the observatory and near the most bright sources of anthropogenic illumination. Snow cover significantly increases the sky brightness due to an increase in the fraction of light reflected by the surface (\cite{snow}). Since we are talking about the southern latitudes of Russian Federation, there may be no permanent snow cover in these regions, and there may be no permanent snow cover in the observatory itself because of strong winds. This increases the scatter of points in the light curves in winter. In addition, we can expect that meteorological parameters such as pressure and the amount of water in the air affect the sky brightness since they change the scattering properties of the air. However, the calculations show that the amount of water has no appreciable effect on the sky brightness in the optical range, and the pressure variation of 25 hPa observed between seasons may lead to variations of $\sim 0.05^m$, which will also be unnoticeable in Fig.~\ref{fig:mag(time)}. 

The general trend of increasing sky brightness can be caused by the combination of solar activity and anthropogenic factors both increasing since 2020.

It is known that during the periods of solar maxima the sky brightness increases, while during the periods of minima~--- decreases. Thus, according to \cite{Benn}, at the La Palma observatory, the sky changed its brightness by $0.4\pm0.1$ magnitude per square second during the solar cycle. 

Both electromagnetic radiation from the Sun and the solar wind affect the brightness of the sky. The former determines the brightness of the atmosphere's own emission (what is called the airglow), and the latter~--- the auroral emission from the upper atmosphere (\cite{Barentine2022}).

We compared the daily average brightness of the moonless night sky with the radio flux at 10.7~cm and the geomagnetic activity index Kp (the data are taken from WDCB\footnote[1]{http://www.wdcb.ru/stp/data.html}). We found high correlation between sky brightness and radio flux ($r=0.76$), and no correlation between sky brightness and the Kp index ($r\approx0.1$). The latter result is rather expected, since at the CMO latitude the contribution of auroral light to the total sky brightness should be small. 

Using the Sky Model Calculator\footnote[2]{https://www.eso.org/observing/etc/bin/gen/form?INS.MODE=swspectr+INS.NAME=SKYCALC}, we calculated the zenith brightness of the winter nighttime moonless sky for the observatory at La Silla (altitude 2400~m) in the Cerro Paranal Advanced Sky Model (\cite{Noll2012}, \cite{Jones2013}). The following parameters were used: a precipitable water volume (PWV) of 3.5~mm, mean monthly radio flux from the Sun of 70~sfu for the minimum and 140~sfu for the maximum of solar activity. As a result, we derived the model $V$ sky brightness at La Silla of $21.97^m$/arcsec$^2$ for the solar activity minimum and $21.69^m$/arcsec$^2$ for the solar activity maximum. Thus, the contribution of solar activity to the value of sky brightness is $\sim 0.3^m$/arcsec$^2$. 

This result cannot be directly transferred to the CMO sky because of the difference in the total sky brightness, whereas the effect of solar activity is just an additive component. We converted the derived contribution of solar activity to the CMO conditions (with a sky brightness of $\sim 20.7^m$/arcsec$^2$ in the $V$ band) and found that the additive effect is $\sim 0.1^m$/arcsec$^2$ during the solar maximum period. Since at the beginning of 2020 the median sky brightness at the CMO was $21.24^m$/arcsec$^2$, and at the end of 2023~-- $\sim 20.5^m$/arcsec$^2$ in the $V$ band, the contribution of solar activity to the change in sky brightness at the CMO in the above time interval is $\sim15$\%. The remaining 85\% is probably due to the anthropogenic factor. 

\section*{Sky spectrum}

\begin{figure*}
\includegraphics[width=\textwidth]{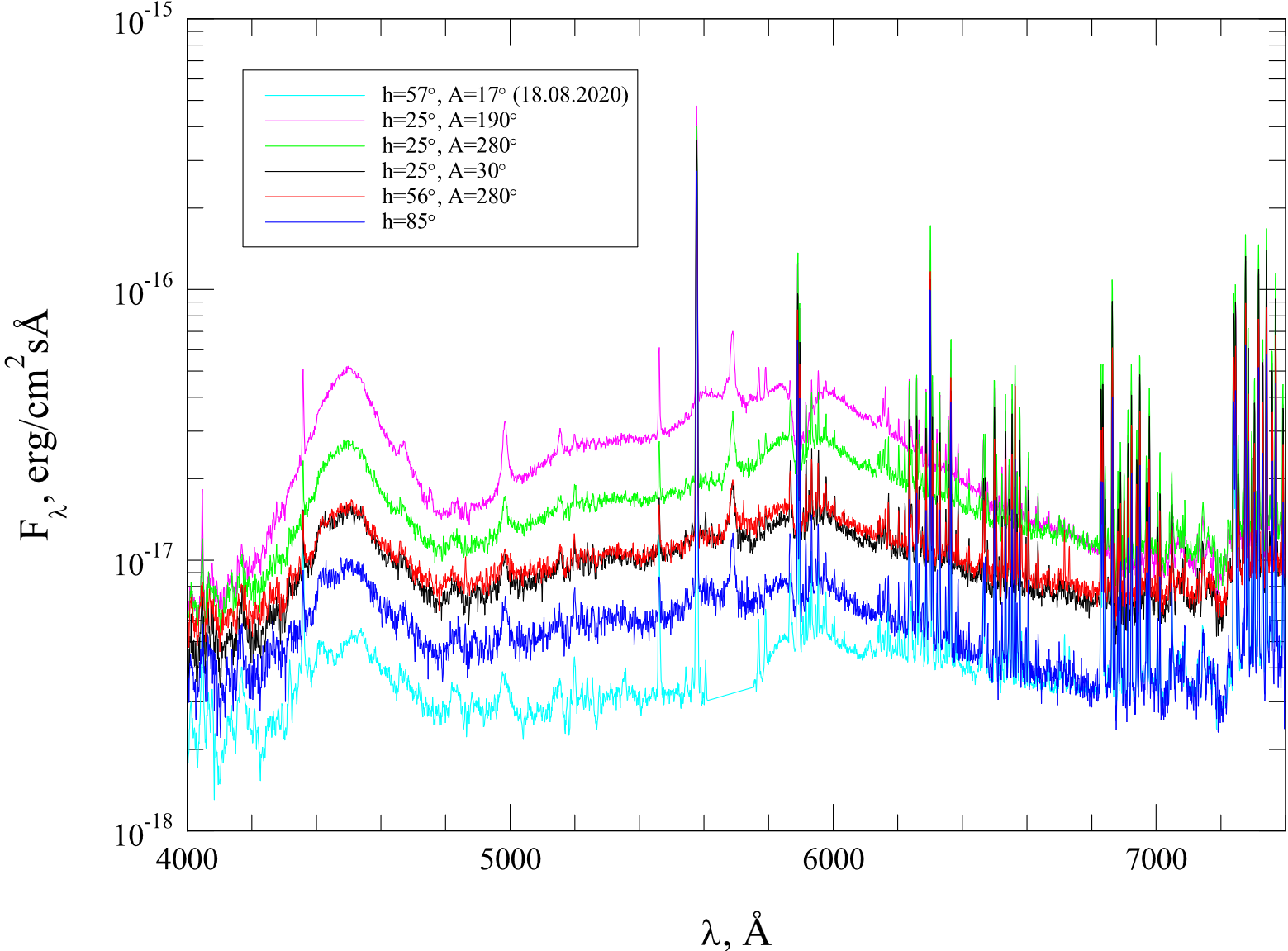} 
\caption{
Spectra of the night moonless sky obtained in the CMO at different altitudes and azimuths in April 2024, and the spectrum obtained on 18.08.2020.
}
\label{fig:sky_spectrum_altaz}
\end{figure*}

In 2020, the Transient Double-beam Spectrograph (TDS) (\cite{Potanin2020}) began operating on the 2.5-m telescope of the CMO. Due to its long slit ($\sim 3^\prime$) and high efficiency, it is possible to simultaneously obtain the spectrum of a program object and the spectrum of the sky with a high signal-to-noise ratio. Fig.~\ref{fig:sky_spectrum_altaz} shows the spectra of the moonless night sky obtained on 11 and 12 April 2024 and, for comparison, the spectrum obtained under similar conditions on 18 August 2020.

When comparing spectra with each other, it should be borne in mind that to the north of the CMO behind the mountain (i.\,e., out of direct line of sight) there is Kislovodsk at a distance of about 20~km, and to the northeast at twice the distance other towns of the Caucasian Mineral Waters resort (Essentuki, Pyatigorsk, etc.) are visible. To the east of the observatory at a distance of several hundred metres, there is a lighted highway running in a ravine about 100~m below the observatory. Therefore, the spectra were recorded at the same height above the horizon ($h=25^\circ$) in the direction of Kislovodsk (azimuth $A=190^\circ$), the highway ($A=280^\circ$), and the dark (so far) Elbrus ($A=30^\circ$). For comparison, the spectra were also obtained near the zenith and above the highway at a height of $56^\circ$.

The sky spectra show a set of emission lines standard for the modern sky (\cite{Ayuga2015}): a "forest"{} of lines of the $OH^-$ molecule in the red part, the [O\,I] ($\lambda\lambda$ 5577, 6300 and 6363~\AA{}), Hg\,I ($\lambda\lambda$ 4047, 4358, 5461, 5770 and 5791~\AA{}), broad Na\,I ($\lambda\lambda$ 4984, 5689, 5890 and 5896~\AA{}) lines. The mercury and sodium lines are anthropogenic in nature and are emitted by luminescent lamps and high-pressure sodium lamps (HPSLs), while the oxygen and hydroxyl lines are produced at altitudes >100~km and are of natural origin. In addition, two broad emission features stand out~--- in the blue part of the spectrum at $\lambda=4500$\AA{} and in the ‘red’{} part at $\lambda=5000...6500$\AA{}. LED lamps (LED, \cite{Ayuga2018}, \cite{Brehm2017}) are responsible for the appearance of the blue feature and the widest part of the ‘red’{} one, which are superimposed on the spectrum of HPSLs (an example of similar spectra is given in Tapia Ayuga et al. 2015). The latter is clearly seen in the wavelength region $\sim 5900$\AA{} in the spectrum obtained in 2020 (when the relative contribution of streetlight LED was smaller). There is a relatively weak blue LED feature in this spectrum, and there is no flux increase characteristic of these modern lamps until almost $\lambda=5700$\AA{}, but the mercury lines 5770 and 5791~\AA{} are well seen, whereas in the 2024 spectra they have less contrast because of the presence of LED emission.

\section*{Discussion}

If there is no anthropogenic illumination, the observed dependence of sky brightness on airmass $Mz$ in some spectral interval can be given by the simple expression

\begin{equation}
F_{obs}(Mz)=(F_s+F_a Mz)\exp{(-\tau_0Mz)},
\label{eq:Fobs}
\end{equation}

where $F_s$ is the galactic (extra-atmospheric) background light, $F_a$ is the brightness of the atmosphere at zenith, and $\tau_0$ is the optical thickness of the atmosphere at zenith. In this expression, we consider that emission originates in the uppermost  atmosphere, while absorption is produced in the lower layers. The introduction of anthropogenic illumination greatly complicates the above expression (see, e.\,g., \cite{Garstang1989}), which lies beyond the scope of this study.

\begin{figure*}
\includegraphics[width=\textwidth]{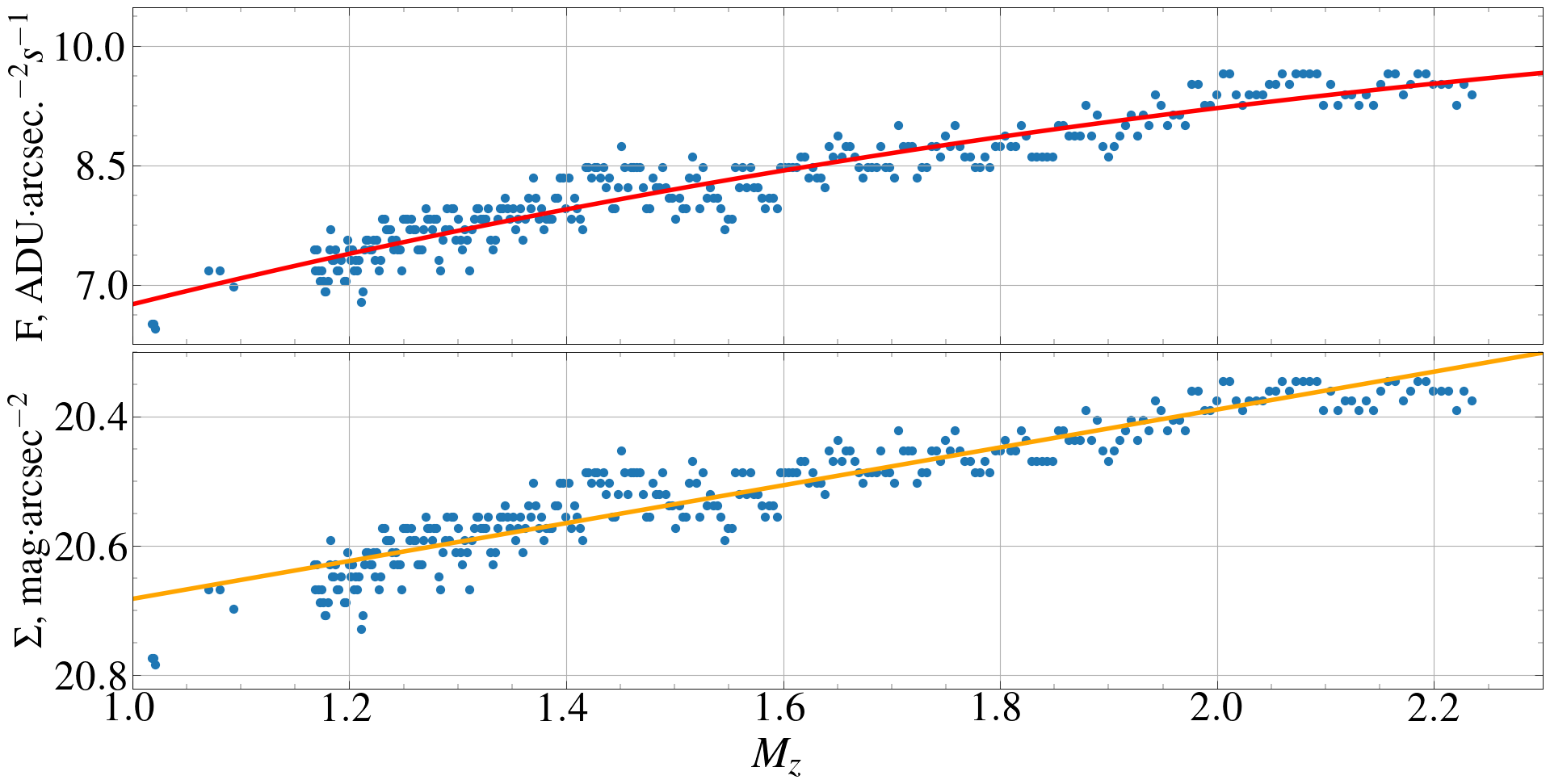} 
\caption{
Variation of the surface sky brightness in the $V$ band with zenith distance during 3.3~h of observations. The red line shows a fit of  $F_{obs}(Mz)$ by the expression \ref{eq:Fobs}, and the orange line is a fit of $mag_{obs}(Mz)$ by linear regression.
}
\label{fig:monitoring}
\end{figure*}

The expression for $F_{obs}(Mz)$ shows that even in the simple case the relation is not linear. The upper panel of Fig.~\ref{fig:monitoring} shows how the sky brightness in units of ADU/(arcsec$^2\cdot$s) changed with $Mz$ during follow-up observations in the $V$ band, conducted on a clear moonless night on 20.09.2023, and by fitting this variation we can determine the parameters $F_s=2.3\pm0.7$, $F_a=6.5\pm1$ and $\tau_0=0.24\pm0.05$ for this date. 

Although the dependence of the logarithm of $F_{obs}$ (i.\,e., the surface brightness in magnitudes) on $Mz$ is not linear, in a sufficiently wide range of $F_s$, $F_a$ and $\tau_0$ values the difference from the straight line fits will be small ($<0.05^m$ at the edges of the working airmass range). This is confirmed by the bottom panel of Fig.~\ref{fig:monitoring} and the appearance of plots in Fig.~\ref{fig:bg(mz)}. Thus, it is possible to use linear regression derived from a large data set to correct the surface brightness to zenith.

\begin{table*}[h]
\caption{
Quartiles of the night sky brightness distribution with the Moon above the horizon and on moonless nights in the $B$, $V$, $R_c$ and $I_c$ photometric bands. Values are given in magnitudes per square arcsecond and refer to zenith. The sky brightness data for 2008--2013 are taken from \cite{Kornilov2016a}. $^*$~--- data for the first month of the year. 
}
\vspace{6pt}\par
\centering
\begin{tabular}{*{10}{c} }
\hline   
\multirow{2}{*}{band} & \multirow{2}{*}{date}&&\multicolumn{3}{c}{without Moon} && \multicolumn{3}{c}{with Moon} \\
&&&25\% & 50 \% & 75 \% && 25\% & 50 \% & 75 \% \\
\hline
\multirow{6}{*}{$B$} & 2008--2013 && 22.31 & 22.10 & 21.84 && 20.87 & 19.73 & 18.96\\ 
& 2019--2020 && 22.07 & 21.90 & 21.76 && 20.24 & 19.12 & 18.53\\
& 2021 && 21.78 & 21.61 & 21.43 && 20.00 & 19.19 & 18.50\\
& 2022 && 21.54 & 21.37 & 21.17 && 20.12 & 18.99 & 18.39\\
& 2023 && 21.44 & 21.31 & 21.16 && 19.95 & 19.00 & 18.40\\
& $2024^*$ && 21.28 & 21.19 & 21.09 && - & - & - \\
\hline
\multirow{6}{*}{$V$} & 2008--2013 && 21.28 & 21.07 & 20.81 && 20.38 & 19.45 & 18.76 \\
& 2019--2020 && 21.40 & 21.27 & 21.14 && 19.98 & 19.08 & 18.62\\
& 2021 && 21.24 & 21.06 & 20.85 && 19.91 & 19.31 & 18.66\\
& 2022 && 20.91 & 20.77 & 20.57 && 20.12 & 19.18 & 18.63\\
& 2023 && 20.74 & 20.64 & 20.51 && 19.77 & 18.85 & 18.30\\
& $2024^*$ && 20.62 & 20.55 & 20.47 && - & - & - \\
\hline
\multirow{5}{*}{$R_c$} & 2019--2020 && 20.70 & 20.58 & 20.44 && 19.99 & 18.96 & 18.47\\
& 2021 && 20.62 & 20.34 & 20.10 && 19.82 & 19.19 & 18.70\\
& 2022 && 20.42 & 20.25 & 20.09 && 19.61 & 19.12 & 18.57\\
& 2023 && 20.25 & 20.16 & 20.02 && 19.67 & 19.05 & 18.17\\
& 2024 && 20.23 & 20.13 & 20.03 && 19.58 & 19.32 & 18.69\\
\hline
\multirow{5}{*}{$I_c$} & 2019--2020 && 19.34 & 19.28 & 19.23 && 18.70 & 18.26 & 18.18\\
& 2021 && 19.49 & 19.28 & 19.11 && 18.93 & 18.63 & 17.98\\
& 2022 && 19.41 & 19.26 & 19.12 && 18.88 & 18.53 & 17.95\\
& 2023 && 19.25 & 19.10 & 18.94 && 18.90 & 18.49 & 17.91\\
& 2024 && 19.36 & 19.19 & 19.01 && 18.89 & 18.51 & 18.01\\
\hline
\end{tabular}
\label{table:coeFFF}
\end{table*}

After correcting all the data to zenith, the median sky brightness was determined, along with the lower and upper quartile values for each photometric band (see Table ~\ref{table:coeFFF}). From the table and Fig.~\ref{fig:mag(time)} we can see that from 2019 to 2024 the brightness of the moonless night sky in the $B$, $V$ and $R_c$ bands increased significantly: the increase was $0.7^m$ in $B$ and $V$, $0.45^m$ in $R_c$. Only in the $I_c$ band the increase was negligible, about $0.1^m$. Similar changes are evidenced by the quartiles. The sky brightness on moonlit nights from 2019 to 2024 changed less, which is explained by the higher level of sky brightness due to scattered Moon light.

With some uncertainty associated with the difference in measuring techniques and bands used, the results obtained for $B$ and $V$ can be compared with the data from \cite{Kornilov2016a} given in Table ~\ref{table:coeFFF} (for the long-wavelength bands such a comparison is impossible due to a large difference between the $R, I$ and $R_c, I_c$ passbands). Note that the data in \cite{Kornilov2016a} were obtained during the phase of increasing solar activity that began after the 2008 minimum. To be able to compare those data with our 2019-2020 data obtained at the solar minimum, we need to apply a correction of $\sim0.1^m$ (see above). The same correction must be applied when comparing the data for 2019 and 2023. The rest of the sky brightness increase is due to the increase in anthropogenic illumination in the neibourhood of the observatory.

In April 2024, we obtained spectra of several sky areas with 15-min intervals (see Fig.~\ref{fig:sky_spectrum_altaz}). Comparing the fluxes in the same telluric lines from fields at different altitudes and putting $F_s=0$ in the expression \ref{eq:Fobs}, we can estimate the absorption at zenith: $\tau_0(5577)\approx0.2$, $\tau_0(7340)\approx=0.1$. In the case of the CMO, these values correspond to clear skies and indicate that spectra are barely affected by local atmospheric anomalies (cloud cover or aerosol scattering). A direct comparison of these spectra shows that the sky brightness at low altitudes can vary 3-4 times, dependent on the proximity of the urban illumination, and confirms the steadily growing input of anthropogenic factors to the sky brightness. And the presence of noticeable broad emission features at zenith and even in the direction to the dark region of the horizon in the southwest, proves that this contribution to the continuum spectrum cannot be ignored in any direction at the CMO.

\section*{Conclusions}

The paper presents the results of processing and analysing tens of thousands of frames obtained from 2019 to 2024 in the $B, V, R_c$ and $I_c$ bands on the 60-cm RC600 telescope installed at the CMO of the SAI MSU. They show that in 5 years the sky brightness in the optical range at the CMO has increased by $0.7^m$, that is, by almost a factor of 2 (see Table ~\ref{table:coeFFF}). As it follows from the expression \ref{eq:snr}, this growth of sky brightness requires a 2-fold increase in integration time for the $B$ and $V$ bands to achieve the previous SNR when observing faint objects. In the red part of the spectrum (the $R_c$ band) the increase was about 50\%, and in the short-wave part of the near-infrared range (the $I_c$ band)~--- only 10\%.

The main contribution to the increase in sky brightness comes from the growth of anthropogenic illumination ($\approx85$\%), caused by the development of cities located to the north and northeast of the observatory at a distance of 20~km or more and the appearance of large greenhouse complexes in their vicinity. The remaining 15\% is the influence of solar activity, which was increasing during the observations analysed in this paper. It should be noted that such a trend is also typical for the worldwide astronomy. In recent years, light pollution has become noticeable even in such previously pristine places as Chilean observatories (La Silla, Gemini (South), etc.~--- see, e.g., \cite{Angeloni2024} and references therein). Not only astronomers, but also medical professionals, ecologists, and wildlife conservationists are alarmed by the increase in artificial lighting over the world. Excessive illumination has turned out to strongly affect living organisms: in humans, the period of circadian rhythms changes, sleep disorders and various health problems occur; in animals, behavioural features change, such as migration (they become disoriented), sleep, search for food, reproduction; also the mortality of nocturnal animal species increases due to weakened vision and reduced ability to avoid predators.

During the period of photometric observations at the CMO, not only the level of sky brightness has changed, but also its spectrum (see Fig.~\ref{fig:sky_spectrum_altaz}). While previously the spectrum was dominated by telluric lines and sodium and mercury lines from street lighting, nowadays LED lamps are the dominant source of illumination in the visible part of the spectrum, giving two broad and bright emission features: the shape of one of them almost coincides with the $B$ passband and the second one occupies the spectrum region from $\sim 5000$\AA{} to $\sim 6200$\AA. At an altitude of $25^\circ$ above the horizon, these emissions are several times brighter in the northern direction (i.\,e., in the direction of nearby cities) than in the southern direction. They are also noticeable at zenith. The descibed changes in sky brightness which are only going to get harder favour observations that are less sensitive to the degree of light pollution - IR photometry and spectroscopy and high-resolution optical spectroscopy.

\section*{Acknowledgments}

The authors are grateful for the support of the Development Programme of Lomonosov Moscow State University (Scientific and Educational School <<Fundamental and Applied Space Research>>).

\textit{Translated by I. Komarova and M. Murlak}

\end{document}